\documentstyle{article}

\input amssym.def
\input amssym.tex

\newcommand{\del}{\partial}
\newcommand{\qed}{\begin{flushright} $\Box$\ \ \ \ \ \
                  \end{flushright}}
\newcommand{\dgleich}{\stackrel{\mbox {\tiny {def}}}{=}}

\newcommand{\real}{{\Bbb R}}
\newcommand{\compl}{{\Bbb C}}

\newtheorem{theorem}{Theorem}[section]

\newtheorem{lemma}[theorem]{Lemma}
\newtheorem{definition}[theorem]{Definition}
\newfont{\largams}{msbm10 scaled 1400} 

\newcommand{\comm}[1]{}

\title{Multicomponent WKB and Quantization}
\author{C. Emmrich, H. R\"omer\\
Fakult\"at f\"ur Physik
\\Universit\"at Freiburg   \\
Hermann-Herder-Stra\ss e 3\\
79104 Freiburg}
\date{}

\begin{document}

\maketitle

\section{Introduction} 
Hamiltonians whose  symbols are not
simply real valued, but matrix or, more generally,  endomorphism valued
 functions appear in many places in physics, examples being the Dirac 
equation, multicomponent wave equations like electrodynamics in media,
and Yang-Mills theories, and the Born-Oppenheimer approximation
in molecular physics.

Whereas the semiclassical and WKB approximation of scalar  systems 
is well understood, this is not the case to the same extent 
for Hamiltonians with matrix valued symbols.  In particular, 
semiclassical states in the scalar case have a nice geometric interpretation
as half densities on Lagrangean submanifolds invariant under the 
Hamiltonian flow, and discrete spectra may be computed using the 
Bohr-Sommerfeld condition, taking into account the Maslov correction
\cite{ma-fe:semiclassical,ba-we:lectures,du:fourier,ho:fourier}.
In the multicomponent case, the analogous structures are not known.
The usual WKB ansatz 
 \[\Psi(x) = a(x,\hbar) \exp(i S(x)/\hbar)\]
 with  $a(x,\hbar) = a_0(x) + a_1(x)\hbar +...$ being a power series 
in $\hbar$ with vector valued coefficients is still possible 
\cite{gu-st:geometric,ka-ma:operators,ma-fe:semiclassical,de:propagation} 
in this situation, but it is in general
defined only locally and leads to equations which don't have an
obvious  geometrical  interpretation. 

The aim of this paper is to give a completely geometric approach 
to this problems, and to reduce the problem ``as far as possible''
to the scalar case. As observed in \cite{li-fl:geometric} the use 
of a  star-product formulation  of quantum mechanics proves 
to be particularly useful in this context. However, whereas
\cite{li-fl:geometric} restrict themselves to the use of the Moyal 
product and thus to the study of trivial
bundles (or local trivializations) over $\real^{2n}$, we will
consider general bundles over arbitrary symplectic manifolds. 
Here, Fedosov's construction \cite{fedosov} will be the 
adequate tool, since it gives an explicit construction for star 
products in this general setting.

\section{WKB on {\largams  R}$^{2n}$ }
In this section, we give a short overview of some well known  results
on WKB for multicomponent systems:
 
Multicomponent WKB deals with equations of the form
 \[ ( \hat{H} -E) \psi = 0 \]
where $\hat{H}$ is an $(N\times N)$-matrix-valued differential operator, 
or equivalently, an $(N \times N)$-matrix
 of differential operators. The WKB-ansatz for a solution is:
\[ \psi= a e^{\frac{i}{\hbar} S}, ~~~ a = a_0 + \hbar a_1 + \ldots \]

At zeroth order, this yields:
\[( H_0(q,d S) -E) a_0 =0,\]
where $H_0$ is the principal symbol of $\hat{H}$, i.e. the zeroth order
part of the symbol of $\hat{H}$. In the scalar case ($N=1$) this simply is 
the  Hamilton-Jacobi-equation for the scalar Hamiltonian  $H_0(q,p)$.

For $N>1$, this equation has two implications: Firstly, $S$ has to be a 
solution  of a Hamilton-Jacobi-equation 
\[ \lambda_\alpha(q,dS) = E ,\]
where $\lambda_\alpha(q,p)  $ is an eigenvalue of $H_0(q,p)$. (Here, we need
the fact that---at least locally---the degeneracies of the eigenvalues do 
not change in order to define the {\em function} $\lambda_\alpha$.)
Secondly, $a_0(q,p)$ has to be an eigenvector of $H_0(q,p)$
with eigenvalue $E$.

Solutions of the Hamilton-Jacobi-equation
define {\em Lagrangean submanifold $\Lambda_\alpha$}
of $\real^{2n}$ and we may 
interpret
\[ a e^{\frac{i}{\hbar} S} \sqrt{d \mu} \]
as a {\em half-density on $\Lambda$} with values in $\real^N$. 
(strictly speaking, it has to be tensorized with a section of
the Maslov bundle).

To study the next order in the WKB approximation, we introduce 
the projector  $ \pi_\alpha^0$ on the $\alpha$th eigenspace of $H_0$. 
We make the assumptions that the  multiplicity of $\lambda_\alpha$
is constant  on a neighborhood of $\Lambda$, 
and that kernel and range of $H-\lambda_\alpha$ are complementary
(which is in particular fulfilled for hermitian $H$).

Then, the first order equation, which is  a ``transport equation'' for 
$u= a \otimes \sqrt{d \mu}$, may be written as \cite{selbstwein}: 
\begin{eqnarray*}
a \otimes {\cal L}_{X_{\lambda_\alpha}} \sqrt{d \mu} 
+ D_{X_{\lambda_\alpha}} a \otimes \sqrt{d \mu}~~~~~~~\hspace{1cm}~~~~~~&& \\
+ \left[- \frac{1}{2} \pi^0_\alpha \{  \pi^0_\alpha, H_0 - \lambda_\alpha 
{\Bbb I}\} \pi^0_\alpha + i \pi^0_\alpha H_1 \pi^0_\alpha \right]  
a \otimes \sqrt{d \mu} &=&0
\end{eqnarray*}

As compared to the scalar case, there are two 
additional terms for $N>1$:
\begin{enumerate}
\item ``Berry term'' for Berry-connection 
\cite{be:quantal,li-fl:geometric,selbstwein}
\[ D a = \sum_\alpha \pi^0_\alpha  d  \pi^0_\alpha a \]
\item
 ``Curvature term'', involves the  curvature $F$ and second fundamental forms
$S, S^*$ of the  Berry-connection. With 
\[
 S \xi \dgleich (1-\pi^0_\alpha) ( d \pi^0_\alpha) \xi, ~~~
S^* \eta \dgleich - \pi^0_\alpha ( d(1-\pi^0_\alpha))\eta 
\]
this term may be rewritten as \cite{selbstwein} 
\[  \lambda_\alpha < \Pi, F> - < \Pi, S^*  \wedge (H_0 - \lambda_\alpha
\pi^0_\alpha) S >,\]
where $\Pi$ is the Poisson tensor.
\end{enumerate}

\section{Star Products and Geometrization}

As mentioned earlier, the star product approach to quantization is particularly
adapted to our problem: Firstly, its structure allows us to deal with the
 expansion in $\hbar$ in a simple way, and secondly, it is the only known 
general quantization scheme which allows the quantization of any symplectic 
manifold, and which is not restricted to  just of a small subclass of 
observables (as, e.g. geometric quantization). The idea  of star product
 quantization is to pull back the operator product, which is defined on
 operators on Hilbert space ${\cal H}$, to an associative, non-commutative
  product on  the functions on phase space, via a suitable symbol calculus
\cite{bayen}: Assuming that we 
 we are given a quantization procedure 
\[ Q:  {\cal C}^\infty(M) \otimes \compl[[\hbar]] \rightarrow 
End {\cal H} \] we may define a star product on ${\cal C}^\infty(M) 
\otimes \compl[[\hbar]]$ by:
\[ a * b = Q^{-1} (Q(a) Q(b)).\]

The simplest example for such a star product is the case
$ M = \real^{2 n } $ and $Q$ being Weyl-ordering, which leads to 
the well known  Weyl-Moyal product. It is explictly given by:
\[ (a \diamond b)(x) = ( e^{ \frac{i \hbar}{2} \omega^{ij} 
\frac{\del}{\del x^i} \otimes \frac{\del}{\del y^j} }
a(x)  b(y) ) |_{x=y} ,\]
where $\omega_{ij} $ are the components of the standard symplectic
 form on $\real^{2 n} $. Although we have only considered scalar 
valued observables so far, the matrix generalization is obvious in this 
simple example: In the matrix product, one simply has to replace the 
ordinary pointwise product of the matrix elements by the Moyal product.

The formal definition is as follows:
\begin{definition}
A star product on a symplectic manifold $M$ is an associatiative
product on $ {\cal C}^\infty(M) \otimes \compl[[\hbar]]$
with
\begin{enumerate}
\item $a * b = a b + O(\hbar)$
\item $ a * b - b*a  = i \hbar \{ a,b\} $
\item $a * 1 = 1 * a = a$
\item $ supp(a * b) \subset supp(a) \cap   supp(b)$
\item $ \overline{a * b}  = \bar{b} * \bar{a} $ 
\end{enumerate}
\end{definition}

Star products exist for every symplectic manifold $M$
 \cite{wil-lec:existence,fedosov}.
Given a star product and a differential operator
 $W = W_0 + \hbar W_1 + \ldots $   
on ${\cal C}^\infty(M) \otimes \compl[[\hbar]]$ with $W 1 = 0$, we may define 
a new star product:
\[ a *_W b = e^{ - \hbar W} ( e^{\hbar W} a * e^{ \hbar W} b). \]
$*$ and $*_W$ are called {\em equivalent}. The importance of this notion is 
that ``physics remains unchanged'', if the  change of $*$ to $*_W$
is accompanied by the application of the  isomorphism $ e^{\hbar W}$ to 
the observables: Heuristically, this means that if we change the ordering 
prescription and at the same time the symbol calculus in such a way that
the operators on Hilbert space remain unchanged, then physics remains 
unchanged.  
We shall exploit this freedom later on.

\section{Fedosov's star product}

In this section, we give a short overview of Fedosov's explicit 
construction of a star product on endomorphism bundles of vector
bundles over a symplectic manifold. The arena is as follows:
We consider an Hermitean vector bundle $V$ over a symplectic manifold $M$.
 Denoting by $\compl[[\hbar]]$ 
the set of formal power series in $\hbar$ with coefficients in $\compl$,
we construct from these data the bundles 
$V[[\hbar]] \dgleich V \otimes  \compl[[\hbar],~
E[[\hbar]] = \cup_{m\in M} End(V_m) \otimes  \compl[[\hbar]]$,
and the algebra 
${\cal A} = \Gamma( E[[\hbar]] \otimes W) $, where  
$W$ denotes the  Weyl-bundle over $M$, the bundle whose  fibres 
$W_q$ are the Weyl algebras over $T_q M$

Finally we consider the algebra: $\Omega(M) \otimes {\cal A}$, with
$\Omega(M)$ being the algebra of differential forms on $M$. 
On this algebra, we define two globally  defined operators 
$\delta, \delta^{-1} : \Omega(M) \otimes {\cal A} \rightarrow
\Omega(M) \otimes {\cal A}$ whose coordinate expressions are given by 
\[\delta \hat{a} = dx^i \wedge \frac{\del}{\del y^i} \hat{a}, ~~
\delta^{-1}  \hat{a} = \frac{1}{p+q} i_{y^i  \frac{\del}{\del x^i}} \hat{a}\]
for a $q$-form $\hat{a}$ with values in the homogeneous polynomials in $y$
of degree $p$. 
Here, $x^i$ are coordinates on $M$, and $y^i$ the induced coordinates
on $T_qM$.   

To construct a unique star product on $E[[\hbar]]$ two additional data 
are required: a  Hermitian connection $\nabla $ on $V$ (inducing
a connection on $E[[\hbar]]$)
and a symplectic connection $\partial_s$ on $M$. From these we can
 construct a connection
$  \partial = 1 \otimes \partial_s + \nabla \otimes 1 $ 
on $E[[\hbar]] \otimes W$. Now, these data given, the main 
step of Fedosov's construction consists in finding 
 by a recursive procedure a  covariant exterior 
derivative 
\[ D = -  \delta + \partial + [ \frac{i}{\hbar} r , \cdot ] \]
with $r \in \Omega^1(M) \otimes {\cal A}$ such that
\begin{enumerate}
\item $D^2 = 0$, i.e., $D$ is a flat connection. 
\item Covariantly constant elements in $  
\Omega_0(M) \otimes {\cal A}$ are in linear one-to-one correspondence
to $\Gamma(E[[\hbar]])$\label{isom}
\item $D$ is a derivation on $\Omega(M) \otimes {\cal A}$ 
 (hence,  covariantly constant sections form a subalgebra)
\end{enumerate}

If we denote the isomorphism in \ref{isom}., which maps $ a \in 
\Gamma E[[\hbar]]$ to its covariantly constant continuation in 
$\Omega_0(M) \otimes {\cal A}$ ,   by $Q$, it is 
given by the solution of
\[ Q(a) =
 a + \delta^{-1} ( \partial Q(a) + [ \frac{i}{\hbar} r, Q(a)] ) ,\]
which may  be solved iteratively.

With these notions, Fedosov's star product is simply given by
 \[ a * b = Q^{-1}(Q(a) \diamond Q(b)),\]
where $\diamond$ is the fibrewise Moyal product.

 Different choices of $\nabla$ (and $\partial_s$)
lead to different  star products. However, the star 
products obtained in this way are equivalent: 

\begin{theorem}\label{th:equiv}
 Let $D^{(1,2)}$ be the  Fedosov connections
associated  to $\nabla^{(1,2)}$,
$\partial^{(1,2)}$. Then there is $A_{(1,2)} \in {\cal A}$ and
\[ U_{(1,2)} = exp_\diamond(\frac{i}{\hbar} A_{(1,2)})\] 
such that
\[ D^{(1)} \hat{a} =0 \Leftrightarrow D^{(2)}( U_{(1,2)} \diamond
\hat{a} \diamond U_{(1,2)}^{-1}) = 0 \]
and \[ \phi(a) \dgleich {Q^{(2)}}^{-1} ( U_{(1,2)} \diamond Q^{(1)}(a)
\diamond U_{(1,2)}^{-1} ) \]
is an isomorphism of the products $*^{(1)}$ and $*^{(2)}$ 
\end{theorem}
\begin{lemma} \label{lemma}
        \[ \nabla^{(1)} - \nabla^{(2)} = O(\hbar^m)\]
\[ \Rightarrow {Q^{(2)}}^{-1} ( U_{(1,2)} \diamond \hat{a}
\diamond U_{(1,2)}^{-1} ) - 
{Q^{(1)}}^{-1} (\hat{a})  = O(\hbar^{m+1})\]
for every $\hat{a}$ with $D^{(1)} \hat{a} = 0 $.
\end{lemma}

\section{Application to WKB}
We are now ready to apply the above techniques to multicomponent 
WKB. Let 
\[ H = H_0 + \hbar H_1 + \ldots\]  be a section of $E[[\hbar]]$,
$ \pi_\alpha^0(x) $ the projection on the  (regular) eigenspace belonging to 
$\lambda_\alpha(x)$. Now, the main problem in WKB, which prevents us 
from naively reducing the problem to a scalar problem on each 
bundle of eigenspaces, is the fact that, due to quantum corrections, 
WKB states are not sections in these bundles, but have higher 
order corrections. In the star product approach this is related 
to the fact that the star commutator of the Hamiltonian with an observable 
commuting with $\pi_\alpha^0(x)$ does not commute with   $\pi_\alpha^0(x)$,
hence a reduction to the bundles of eigenspaces is not possible. 
Thus, we have to find a kind of ``quantum diagonalization procedure''.

The formal reason for this problem in Fedosov's approach to 
star products is the fact that the 
projections are not covariantly constant under the hermitian connection
which defines the star product.
Our strategy for solving this problem is to 
use the freedom to change $\nabla$ when  applying the 
corresponding  star product isomorphism at the same time
to preserve physics. This is possible indeed, as expressed by
the following theorem:

\begin{theorem}
There exists a formally orthogonal decomposition $V[[\hbar]] = 
\oplus_\alpha V_\alpha,$ $ \dim V_\alpha = m_\alpha$
with corresponding quantum projections $\pi_\alpha^{(\infty)}$ and 
a hermitian connection $\nabla^{(\infty)}$ with
\[ \nabla^{(\infty)} \pi_\alpha^{(\infty)} = 0 \]
such that the corrected Hamiltonian (obtained by 
applying the corresponding isomorphism $\phi^{(\infty)}$):
\[ H^{(\infty)} = \phi^{(\infty)}( H) \]
preserves the decomposition:
\[ [ H^{(\infty)}, \pi_\alpha^{(\infty)} ]_{*_\infty} = 0 ~~ \forall \alpha ,\]
where $*_\infty$ is the Fedosov star product defined by 
$\nabla^{(\infty)}$
\end{theorem}

\noindent {\em Proof:}

The proof is by induction: 
To start we set $ \pi_\alpha^{(1)} = \pi_\alpha^{(0)}$ and define 
$ \nabla^{(1)}$ by $ \nabla^{(1)} \psi
= \sum_\alpha \pi_\alpha^{(1)} \nabla \pi_\alpha^{(1)} \psi$.
With $ H^{(1)} = \phi^{(1)}(H)$, we have
\[ \nabla^{(1)} \pi_\alpha^{(1)} = 0, ~~~ [ \pi^{(1)}, H^{(1)}]_{*_1} = 
O(\hbar).\]

Now assume we have found
$ \pi_\alpha^{(k)},  
 \nabla^{(k)} 
= \sum_\alpha \pi_\alpha^{(k)} \nabla \pi_\alpha^{(k)} $,
$ H^{(k)} = \phi^{(k)}(H)$
such that 
\[ \nabla^{(k)} \pi_\alpha^{(k)} = 0, ~~~ [ \pi^{(k)}, H^{(k)}]_{*_k} = 
O(\hbar^k)\]

 We construct $\pi_\alpha^{(k+1)} $ with
\[ [ \pi_\alpha^{(k+1)},H^{(k)}]_{*_k} = O(\hbar^{(k+1)}) \]
using the ansatz 
\[\pi_\alpha^{(k+1)} = e^{i \hbar^k A} *_k  \pi_\alpha^{(k)} *_k
e^{-i \hbar^k A}, \]
which has a solution
\[ A = \sum_{\alpha\neq \beta} \frac{ 
  \pi_\alpha^{(k)} W   \pi_\beta^{(k)}}{ \lambda_\alpha -
\lambda_\beta}  \]
where $W$ is defined by
\[ H^{(k)} = \sum_\alpha   \pi_\alpha^{(k)} *_k H^{(k)} *_k
  \pi_\alpha^{(k)} + \hbar^k W. \]

Now, setting $\nabla^{(k+1)} 
= \sum_\alpha \pi_\alpha^{(k+1)} \nabla \pi_\alpha^{(k+1)} \psi$
and using lemma \ref{lemma},
the theorem follows. 
\qed

\section{Compatibility}
Now, we are ready to address the issue of compatibility of
observables, i.e., the question which observables 
preserve the quantum decomposition found above. The results
may be considered generalizations of results found in 
\cite{co:dirac} for the Dirac equation. In physics language, 
we find an answer to the question which observables are ``slow''
in the sense that their time evolution is analytic in $\hbar$, and 
hence does not depend on inverse powers of $\hbar$.

Take $a,b \in \Gamma(E[[\hbar]])$ such that
\[ [  \pi_\alpha^{(\infty)},a]_{*_\infty} = 
[  \pi_\alpha^{(\infty)},b]_{*_\infty}
=0\]
then 
\[ [   \pi_\alpha^{(\infty)}, a *_\infty b]_{*_\infty} = 0 \]
Compatible elements form a $*_\infty$-subalgebra 
$\tilde{\cal O}$. In particular, for $b=H^{(\infty)}$
we see that the star commutator of observables in 
$\tilde{\cal O}$ with the corrected Hamiltonian is
in $\tilde{\cal O}$ again. 

\begin{theorem}
Let $\phi^{(\infty)} $ be the isomorphism between the star products
$*$ and $*_\infty$. 
For every $a $ in the subalgebra ${\cal O} = {\phi^{\infty}}^{-1}
(\tilde{\cal{O}})$ the Heisenberg evolution equation
\[ \dot{a} = \frac{1}{i \hbar}[ H,a]_* \]
is a well defined differential equation in ${\cal O}$
(i.e., there are no inverse powers of $\hbar$ on the
right hand side).
\end{theorem}

\section{$O(h\hspace{-0.4em}
\protect\rule[1.2ex]{0.8ex}{0.1ex})$-correction}
In the formalism developped so far, it is straightforward 
to compute the correction of order $\hbar$: We have
\[ H^{(\infty)} =    \sum_\alpha \pi_\alpha^{(2)}*_1 H^{(1)}*_1
\pi_\alpha^{(2)} + O(\hbar^2) \]
\[ = e^{i \hbar A} * \pi_\alpha^{(0)} * H^{(1)} * \pi_\alpha^{(0)} * 
  e^{-i \hbar A} + O(\hbar^2)\]
A straightforward    calculation, using the explicit 
formula for $U$  leads to \cite{emmroem}
\[ \pi_\alpha^{(0)} *_1 H^{(1)} *_1 \pi_\alpha^{(0)} = 
\pi_\alpha^{(0)} H^{(0)} \pi_\alpha^{(0)} + \frac{i \hbar}{2}
\omega^{il} \sum_\gamma  \pi_\alpha^{(0)} ( \nabla_i \pi_\gamma^0)
\nabla_l \pi_\alpha^{(0)} \]

With the curvature of the ``Berry connection''
\[ F^B \psi =  \sum_\alpha  \pi_\alpha^{(0)} \circ  \nabla 
\circ \pi_\alpha^{(0)}
\nabla  \pi_\alpha^{(0)} \psi 
= \sum_\alpha  \pi_\alpha^{(0)} F^0 \pi_\alpha^{(0)}
+ \sum_\alpha 
  \pi_\alpha^{(0)} ( \nabla \pi_\gamma^0) (\nabla \pi_\alpha^{(0)})
\]
and the  ``Second fundamental form'' 
\[ S^{\beta \alpha}(X) \psi = (
  \pi_\beta^{(0)}\circ   \nabla _X \circ \pi_\alpha^{(0)}) \psi ~~~ (\beta
\neq \alpha) \]
we finally get:
\[  \pi_\alpha^{(0)}* H^{(1)}  * \pi_\alpha^{(0)} =
\frac{i \hbar}{2} \lambda_\alpha <\Pi, \pi_\alpha^0 (F^B- F^0) \pi_\alpha^0>
+ \frac{i \hbar}{2} \sum_{\gamma \neq \alpha} \lambda_\gamma
< \Pi, S^{\gamma \alpha} S^{\alpha \gamma}>.
\]

This result is a generalization of our results in $\cite{selbstwein}$:
There, the starting point was the Moyal product on $\real^{2n}$,
which is a Fedosov star product for a flat connection. Hence, the 
curvature  $F^0$ of $\nabla$ was missing there.

\section{Change of compatible connection}

In the constructions above we have made explicit use
of Berry type connections. Nevertheless, we are not really 
forced to do so, but it is just a convenient choice for
the proof. Hence, the question arises whether the correction 
terms computed above are just a consequence of the choice
of our connection, or whether they are really ``physical''.

To answer this question, take another
 compatible connection $\tilde{\nabla}$.
With 
\[ \nabla^{(1)} - \tilde{\nabla}^{(1)} = \Delta \Gamma\]
we obtain
\[ \pi_\alpha^{0} * \tilde{H}^{(1)} * \pi_\alpha^{0} = 
 \pi_\alpha^{0}* H^{(1)}*  \pi_\alpha^{0} 
+ i \hbar  \pi_\alpha^{0} \underbrace{\Delta\Gamma(X_{\lambda_\alpha})
}_{\mbox{``Berry phase'' term}} \pi_\alpha^{0} \]

Hence, the only difference is the appearence of an additional 
Berry phase, which was so far hidden in the explicit use 
of a non-flat connection. However, the curvature term remains
and it still is constructed from the curvature of the Berry 
connection, {\em not} from the curvature of the chosen connection. 

{\em The correction terms involving the curvature and the second 
fundamental form of the Berry connection have intrisic meaning.}

\section{Problems}
In this section, we just mention a few open problems which 
are discussed in more detail in \cite{emmroem}.
\begin{enumerate}
\item Nontrivial holonomy of compatible connections may give an
obstruction  to  the existence of WKB states:
only projectors exist because of a $U(n)$ holonomy.
Solution are known  only for
 $m_\alpha=1$(no degeneracy) or $\dim M=2$ \cite{ka:global,emmroem}
\item  A Maslov-correction has to be implemented. This 
is straightforward if the problem can really be reduced 
to a scalar problem, e.g. for $m_\alpha=1$.
\item Level crossings appear in many physical applications. 
These have been excluded by our regularity conditions. 
Nevertheless, our approach is still applicable to the 
open submanifold where the degeneracies are constant. The
level crossing points have to be studied in a second step. 
\end{enumerate}


\begin{thebibliography}{99}
\bibitem{ba-we:lectures}
Bates,~S.M., and Weinstein,~A.,  Lectures on the Geometry of Quantization,
{\em Berkeley Mathematics Lecture Notes }
\bibitem {bayen}
         {F. Bayen, M. Flato, C. Fronsdal, A. Lichnerowicz,
            D. Sternheimer:}
         {Deformation Theory and Quantization.}
         {\it Annals of Physics} {\bf 111} (1978), part I: 61-110,
         part II: 111-151.
\bibitem{be:quantal}
Berry,~M.V., Quantal phase factors accompanying adiabatic changes,
{\em Proc. R. Soc. London A} {\bf 392} (1984), 45-57.
\bibitem{co:dirac}
Cordes,~H.O., A pseudo-algebra of observables for the Dirac equation,
{\em Manuscripta Math.} {\bf 45} (1983), 77-105.
\bibitem{de:propagation}
Dencker,~N., On the propagation of polarization
sets for systems of real principal type, {\em
J. Funct. Anal.} {\bf 46} (1982), 351-372.
\bibitem {wil-lec:existence} {M. DeWilde, P.B.A. Lecomte:}
         {Existence of star-products and of formal deformations
         of the Poisson Lie Algebra of arbitrary symplectic manifolds.}
         {\it Lett. Math. Phys.} {\bf 7} (1983), 487-496.
\bibitem{du:fourier}
  Duistermaat,~J. J., Fourier Integral Operators,{\em Courant
  Institute}, NYU, New York, 1973.
\bibitem{fedosov}  B. Fedosov, 
          A Simple Geometrical Construction of Deformation Quantization,
          J. of Diff. Geom. {\bf 40} (1994), 213-238.         
\bibitem{gu-st:geometric}
  Guillemin,~V., and Sternberg,~S., Geometric Asymptotics, 
  {\em Math. Surveys} {\bf 14}, Amer. Math. Soc., Providence, 1977.
\bibitem{ho:fourier} 
H\"ormander,~L., Fourier Integral Operators I., {\em Acta Math.}
{\bf 127} (1971) 79-183.
\bibitem{ka:global}
Karasev,~M.V., New global asymptotics and anomalies for the problem of
quantization of the adiabatic invariant,  {\em Funct. Anal. Appl.} {\bf 24}
(1990), 104-114.
\bibitem{ka-ma:operators} 
Karasev,~M.V., and Maslov,~V.P., Operators with general commutation
relations and their applications.  I.  Unitary-nonlinear operator
equations, {\em J. Sov.  Math.} {\bf 15} (1981), 273-368.
\bibitem{li-fl:geometric}
Littlejohn,~R.G., and Flynn,~W.G., Geometric phases in the asymptotic
theory of coupled wave equations,
{\em Phys. Rev.} {\bf A44} (1991), 5239-5256.
\bibitem{ma-fe:semiclassical}
Maslov,~V.P., and  Fedoriuk,~M.V.,
{Semi-classical Approximation in Quantum Mechanics},
D. Reidel, Dordrecht, 1981.
\bibitem{selbstwein} Emmrich,~C., and Weinstein,~A.,
Geometry of the transport equation in multicomponent WKB approximation,
 to appear in {\em Comm. Math. Phys.}
\bibitem{emmroem} Emmrich,~C., and R\"omer,~H., 
{Multicomponent WKB on arbitrary symplectic manifolds: 
A star product approach}, in preparation 
\end{thebibliography}
\end{document}